\documentclass[prb,aps,epsf,twocolumn,showpacs,superscriptaddress]{revtex4}
\usepackage[dvips]{graphicx}
\usepackage{latexsym}
\usepackage{amsmath}
\begin{document}

\title{Critical fermi surfaces and non-fermi liquid metals}

\author{T. Senthil}
\affiliation{ Department of Physics, Massachusetts Institute of
Technology, Cambridge, Massachusetts 02139}

\date{\today}
\begin{abstract}
At certain quantum critical points in metals an entire Fermi surface
may disappear. A crucial question is the nature of the electronic excitations at the critical point. Here we provide arguments showing that at such quantum critical points the Fermi surface remains sharply defined even though the Landau quasiparticle
is absent. The presence of such a critical Fermi surface has a number of consequences for the universal phenomena near the quantum critical point which are discussed. In particular the structure of scaling of the universal
critical singularities can be significantly modified from more familiar criticality. Scaling hypotheses appropriate to a critical fermi surface are proposed. Implications for experiments on heavy fermion critical points are discussed. Various phenomena in the normal state of the cuprates are also examined from this perspective. We suggest that a phase transition that involves a dramatic reconstruction of the Fermi surface might underlie a number of strange observations in the metallic states above the superconducting dome.

\end{abstract}
\newcommand{\be}{\begin{equation}}
\newcommand{\ee}{\end{equation}}
\maketitle

\section{Introduction}
Several recent developments show that at certain quantum phase transitions in metals an entire Fermi surface
may disappear.
If such transitions are second order it may be expected that the corresponding ground states are non-fermi liquid metals. In light of the long standing mysteries associated with the theory of non-Fermi liquid metals it is thus important to explore the nature of such quantum critical points associated with the disappearance of a
Fermi surface.
In this paper we first argue that the ground state of such a quantum critical point will be characterized by a sharp Fermi surface even though the Landau quasiparticle may not be well defined. We will provide a general scaling theory of such a quantum phase transition focusing in particular on the critical excitations associated with the disappearing Fermi surface.

\begin{figure}
\includegraphics[width=8cm]{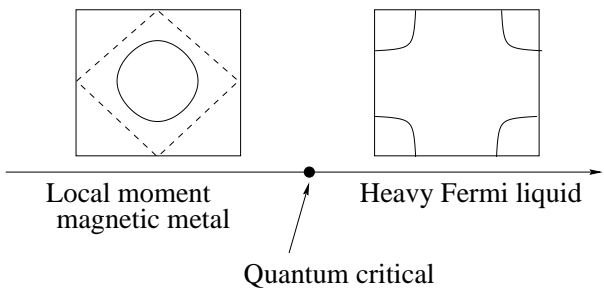}
\caption{Possible schematic zero temperature phase diagram showing the onset of magnetism in a heavy fermion metal. The magnetic phase is a `local moment magnetic metal' where the local moments are not part of the Fermi surface unlike in the heavy Fermi liquid. The Fermi surface needs to reconstruct across such a quantum phase transition.} \label{hfpdia}
\end{figure}

\begin{figure}
\includegraphics[width=8cm]{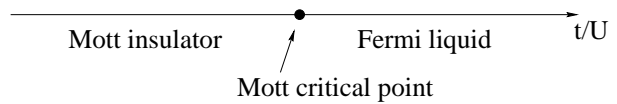}
\caption{Possible schematic zero temperature phase diagram for a half-filled single band repulsive Hubbard model on a non-bipartite lattice. $U$ is the Hubbard interaction strength and $t$ is the hopping amplitude. The Fermi surface of the metal needs to disappear at the Mott transition.} \label{mottpdia}
\end{figure}

\begin{figure}
\includegraphics[width=8cm]{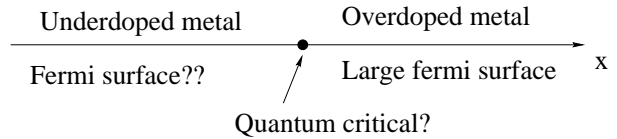}
\caption{Possible schematic zero temperature phase diagram for the cuprate materials showing the evolution of the `underlying normal' ground state as a function of doping.  The large Fermi surface of the overdoped metal is presumed to disappear at a critical doping $x_c$ to an underdoped metal with a qualitatively different `small' Fermi surface.} \label{hitcpdia}
\end{figure}

Let us first list specific situations to which the considerations of this paper are pertinent.
 \begin{enumerate}
 \item
 \underline{Heavy fermion criticality}

 The onset of antiferromagnetism in heavy electron materials such as $CeCu_{6-x}Au_x$, $CePd_2Si_2$, $YbRh_2Si_2$, or the ``$115$" compounds ($CeCoIn_5, CeRhIn_5$) is known to be accompanied by the breakdown of Fermi liquid theory\cite{hfspht,hftrans,hfrevs,hfcscal}.
There is good evidence that the  corresponding quantum phase transition is second order. The Neel temperature appears to go to zero continuously. Several characteristic singularities are seen, for instance in the specific heat\cite{hfspht}. Finally scaling has been demonstrated in both thermodynamic quantities like the specific heat\cite{hfcscal} and also in the dynamical spin correlations at the critical point\cite{schroder}. Theoretically this behavior is inconsistent with the `standard' theory of the onset of magnetism in a metallic environment\cite{Hertz} (due to Moriya, Hertz, Millis, and others). This has prompted the idea that the magnetic transition may be accompanied by a fundamental change of the electronic structure associated with the possible breakdown of Kondo screening\cite{cpsr,qimiao,svsrev}. Fluctuations associated with this change in electronic structure might then underlie the observed non-fermi liquid quantum criticality. Within this thinking a drastic change of the topology of the Fermi surface might be expected across the transition (see Fig. \ref{hfpdia}).

Remarkably recent experiments indicate that the Fermi surface may indeed reconstruct rather dramatically across the quantum phase transition. One piece of evidence is the evolution of the Hall coefficient in $YbRh_2Si_2$ across the transition\cite{Silke}. Other evidence comes from deHaas-van Alphen studies of $CeRhIn_5$ as a function of pressure\cite{onuki}. This material is an antiferromagnetic metal at ambient pressure. The antiferromagnetism is lost at a critical pressure of 2.35 GPa. Remarkably the dHvA frequencies jump at exactly this critical pressure. Furthermore the effective mass on various Fermi surface sheets seems to diverge at the critical pressure. It appears therefore that entire sheets of the Fermi surface are disappearing as the quantum critical point is approached.

\item
\underline{Mott criticality}

A rather different quantum phase transition is also associated with the disappearance of a Fermi surface - namely the Mott metal-insulator transition at fixed filling in a one band Hubbard model on a non-bipartite lattice (Fig. \ref{mottpdia}). Recent experiments on the layered triangular lattice organic material $\kappa-(ET)_2Cu_2(CN)_3$ (believed to be described correctly by a single band Hubbard model) have begun to probe such a transition on an isotropic triangular lattice\cite{kanoda1}. There are again indications that the transition may be second order\cite{kanoda2} even though many more future studies are needed to establish this.
If a second order Mott transition is indeed possible then again the entire Fermi surface of the metal needs to disappear in a continuous way at the transition point.

\item
\underline{Cuprate metals}

A third example is provided by the cuprate high temperature superconducting materials. In the overdoped side it is very likely that the `underlying normal' ground state ({\em i.e} the ground state in the absence of superconductivity) is a Fermi liquid with a large Fermi surface satisfying Luttinger's theorem\cite{overdpfs}. In the underdoped side on the other hand it is hardly clear that this continues to be the case. Recent experiments show that when the superconductivity of an underdoped cuprate is suppressed by a field a metallic state results\cite{underdposc}. The nature of that metallic state is not settled - however it seems clear that it is not smoothly connected to the overdoped Fermi liquid. A number of possible states have been described in the literature which have small hole pockets. Motivated by this we will here make the assumption that there is a critical doping $x_c$ such that the large Fermi surface Fermi liquid is stable only for $x > x_c$ (see Fig. \ref{hitcpdia}). As the quantum phase transition at $x_c$ is approached from the overdoped side the large Fermi surface disappears. On decreasing $x$ below $x_c$ the underdoped metallic state results. If the phase transition at $x_c$ is second order, then the ground state at $x_c$ will have a critical Fermi surface and will describe a strange metal. We will show how the general scaling theory of such transitions developed in this paper  might accomodate a number of the
mysterious normal state phenomena in the cuprates.

\end{enumerate}

How can a Fermi surface disappear continuously at any of the phase transitions discussed above? One route that has been discussed extensively\cite{BrRice} is that the quasiparticle residue $Z$ vanishes continuously as the transition is approached. The crucial question then is the fate of the Fermi surface at the critical point when $Z$ has just gone to zero. Here we first provide arguments showing that at the critical point the Fermi surface remains sharply defined despite the vanishing $Z$.

It is instructive to first think about the Mott critical point discussed above. The spectrum of single particle excitations at zero temperature is conveniently described through the
electron spectral function $A(\vec K, \omega)$. In the Fermi liquid phase this has delta function quasiparticle peaks at the Fermi surface. In the Mott insulator  $A(\vec K, \omega) = 0$ for $\omega < \Delta(\vec K)$ for any fixed $\vec K$. The $\Delta(\vec K)$ is the single particle gap at momentum $\vec K$ and is sharply defined. Clearly on approaching the Mott critical point this gap has to close. With a continuous transition to the Fermi liquid we expect that this gap will go to zero at all $\vec K$ points that correspond to the Fermi surface of the metal. The criticality is then associated with gapless single particle excitations through out the Fermi surface. The Fermi surface will thus be sharp at the critical point. However as $Z$ is also zero we have no Landau quasiparticle.
We are therefore very naturally lead to the possibility that the quantum critical point is characterized by a sharp Fermi surface but with no quasiparticle peak in the spectral function. We will dub this a `critical Fermi surface'.

Similar reasoning (for a preliminary discussion see Ref. \onlinecite{tsannals}) in the heavy fermion context again leads to the possibility that the quantum critical metal  has a critical Fermi surface. Specifically on approaching the transition from the heavy Fermi liquid side
the large Fermi surface needs to disappear. Likewise the Fermi surface of the antiferromagnetic metal also needs to disappear when the transition is approached from the antiferromagnetic side. This can happen by the quasiparticle residues vanishing on both Fermi surfaces on approaching the critical point\cite{qimiaosces,tsannals}. Right at the critical point
the delta function quasiparticle peak is expected to be replaced by a universal power law singularity on both Fermi surfaces. Thus the quantum critical state may again be expected to have sharp critical fermi surfaces but no sharp quasiparticle.

In the context of the cuprate metals, the same considerations also lead us to the concept of critical Fermi surfaces at the critical doping $x_c$ described above. The possibility of a state with a sharp Fermi surface but no sharp quasiparticle was advocated by Anderson
as a description of the optimally doped cuprates\cite{pwa}. A sharp Fermi surface with no Landau quasiparticle is also a crucial ingredient of the marginal Fermi liquid phenomenology proposed by Varma et. al\cite{Varma} for optimally doped cuprates. We have argued above that a natural realization of such  states is at a quantum critical point where a Fermi surface disappears. Toward the end of this paper we examine phenomena in the normal state of the cuprates from the perspective taken in this paper.

\begin{figure}
\includegraphics[width=8cm]{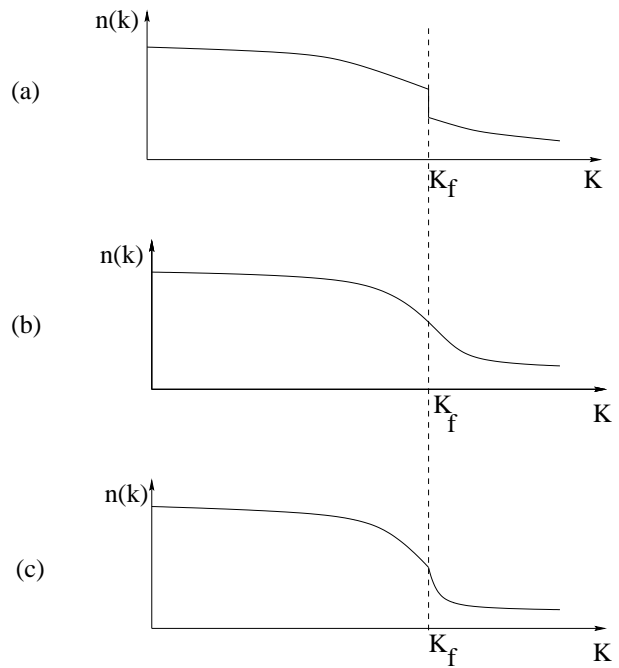}
\caption{Evolution of the ground state momentum distribution $n(k)$ across a second order phase transition where the Fermi surface disappears, such as the Mott transition of Fig. \ref{mottpdia}. (a) $n(k)$ in the Fermi liquid with a discontinuity $Z$ at the Fermi surface.  (b) $n(k)$ in the Mott insulator which is smooth as a function of $k$. (c) $n(k)$ at the critical point - the discontinuity of (a) has just vanished and is replaced by a kink singularity.} \label{nkqcrit}
\end{figure}

It is also useful to think about the evolution of the ground state momentum distribution through a phase transition where a Fermi surface disappears. This is depicted in Fig. \ref{nkqcrit}. In the phase where the Fermi surface is present the momentum distribution has a jump discontinuity $Z$ at the location of the Fermi surface. In the other phase the momentum distribution will be smooth at the same location. A second order transition between the two phases requires that the jump vanish on approaching the transition from the side with the Fermi surface present. Right at the critical point it is natural then to expect that the jump is replaced by a kink singularity.
Thus the Fermi surface will still be sharply defined at the critical point even though the Landau quasiparticle is not.

The
existence of a sharp `critical Fermi surface' may be expected to profoundly influence the structure of the universal singularities at the quantum critical point. Criticality is in general associated with the phenomenon of scaling in various physical quantities. The structure of
scaling phenomena at bosonic quantum critical points is well understood\cite{subirbook}. However it is clearly inappropriate to expect the exact same scaling at fermionic quantum critical points with a gapless Fermi surface.
In the rest of this paper we formulate scaling hypotheses for non-fermi liquid states with `critical' Fermi surfaces. These hypotheses are natural generalizations of the ones for familiar bosonic criticality to situations with a Fermi surface. We will show that the presence of a Fermi surface crucially impacts the universal singular behavior of almost all physical properties and leads to striking differences from bosonic quantum critical points.

The rest of the paper is organized as follows. We begin in Section \ref{elssclhyp} by formulating scaling hypotheses for the single particle spectral function associated with a critical Fermi surface. We point out that in principle the exponents characterizing the universal singularities may depend on angular position on the Fermi surface. Exponent inequalities are derived. Next in Section \ref{othersclhyp} we formulate scaling hypotheses for various thermodynamics and two-particle correlators. The thermodynamics is considered first in subsection \ref{thermo}. Several possible distinct scaling models are argued to exist and the nature of the corresponding singularities are discussed. Then in subsection \ref{2kf} we consider the singularities in two particle correlators such as the spin density. We propose that due to the critical Fermi surface there will be a sharp ``$2K_f$" surface at which such two particle correlators will have universal singularities. A scaling ansatz for the universal behavior near this critical $2K_f$ surface is formulated. In Section \ref{fermiarcs} we consider the remarkable consequences of angle dependent exponents for the finite temperature crossovers near the quantum critical point. In particular the crossover will include an extended regime in temperature of a metal with $T$-dependent gapless Fermi arcs. Motivated by this observation we provide, in Section \ref{cuprates} a tentative application of the scaling ideas of this paper to normal state phenomena in the cuprates. We explore the possibility that a quantum critical point associated with a dramatic reconstruction of the Fermi surface might underlie many of the mysterious normal state phenomena. We show that the scaling perspective developed in this paper provides a very appealing framework for thinking about these phenomena. Next in Section \ref{slave} we provide some theoretical evidence for the existence of a critical fermi surface by a simple slave particle mean field calculation for a Mott transition in a non-bipartite lattice. Finally in Section \ref{disc} we discuss several questions raised by these ideas, and their implications for various experiments.

\section{Scaling hypotheses for the single particle spectral function}
\label{elssclhyp}
In a Fermi liquid with Landau quasiparticles at a Fermi surface the electron spectral function has the asymptotic form
\begin{equation}
\label{elsfn_lfl}
A(\vec K, \omega) = \frac{1}{\pi}\frac{Z\gamma}{\left(\omega - v_F k_{\|}\right)^2 + \gamma^2}
\end{equation}
where $k_\|$ is the shortest deviation of the momentum $\vec K$ from the Fermi surface ({\em i.e} distance in momentum space parallel to a normal to the surface at the point of closest approach). $Z$ is the quasiparticle residue and $v_F$ is the Fermi velocity. The quasiparticle decay rate $\gamma \sim \omega^2$. This may thus be approximated by a delta function
\begin{equation}
A(\vec K, \omega) = Z \delta (\omega - v_F k_{\|})
\end{equation}
This delta function form is asymptotically exact in the limit of small $\omega, k_\|$. We note that in general in a lattice system $Z$ and $v_F$ may depend on position on the Fermi surface {\em i.e} on $\hat{K_F}$.

On approaching the critical point where this Fermi surface disappears $Z$ will go to zero. Right at the critical point $A(\vec K, \omega)$ is expected to have a universal singular dependence on $\omega$ and $k_\|$ when these are both small. It is natural to expect that this singular structure is described by a scaling ansatz of the form
\begin{equation}
\label{elsfn_cfs}
A(\vec K, \omega) \sim \frac{c_0}{|\omega|^{\frac{\alpha}{z}}} F_0\left(\frac{c_1 \omega}{k_{\|}^z}\right)
\end{equation}
Here the critical exponents $z,\alpha$ are universal as is the scaling function $F_0$. The $c_{0,1}$ are non-universal constants. This equation is our primary scaling hypotheses and is an obvious generalization of the well-known scaling structure of bosonic critical points to situations with a critical fermi surface. The scaling ansatz is readily generalized to non-zero temperatures and takes the form
\begin{equation}
\label{elsfn_cfsT}
A(\vec K, \omega, T) \sim \frac{c_0}{|\omega|^{\frac{\alpha}{z}}} F_T\left(\frac{c_1 \omega}{k_{\|}^z}, \frac{\omega}{T}\right)
\end{equation}

A new possibility that exists with a critical Fermi surface is that the critical exponents $z, \alpha$ may depend on position on the Fermi surface, {\em i.e} in general
\begin{eqnarray}
z & = & z(\hat{K_F}) \\
\alpha & = & \alpha(\hat{K_F})
\end{eqnarray}
This angle dependence will be restricted by lattice symmetries. For instance on the triangular lattice if we parametrize the position on the Fermi surface by an angle $\theta$ we have $z(\theta) = z(\theta + \frac{\pi}{3})$ and similarly for $\alpha$.

General considerations allow us to obtain some important restrictions on the critical exponents $z$ and $\alpha$. First consider the electron momentum occupation function $n(\vec K) = <c^\dagger_K c_K> $ at zero temperature in the vicinity of the Fermi surface. This is obtained from $A(\vec K, \omega)$ through
\begin{equation}
n(\vec K) = \int_{-\infty}^0 d\omega A(\vec K, \omega)
\end{equation}
In the Fermi liquid when the Landau quasiparticle is well-defined $n(\vec K)$ has a jump discontinuity at the Fermi surface. The jump $Z$ goes to zero at the critical point and will be replaced by a universal kink singularity. The scaling ansatz above readily allows obtaining the form of this singularity. We find (for the singular part)
\begin{equation}
n(\vec K) \sim |k_{\|}|^{z - \alpha}
\end{equation}
As in an electronic system $n(\vec K)$ must clearly always be bounded we have the inequality
\begin{equation}
z(\hat{K_F}) \geq \alpha(\hat{K_F}).
\end{equation}
This inequality must hold at every point on the Fermi surface. Other inequalities will be derived below.

A scaling form for the tunneling density of states at low frequency and temperature
\begin{equation}
N(\omega, T) = \int \frac{d^d\vec K}{(2\pi)^d} A(\vec K, \omega)
\end{equation}
is readily derived. The singular part $N_s(\omega,T)$ of $N(\omega, T)$ is given by
\begin{eqnarray}
N_s(\omega, T) & \sim & \int d^{d-1} \vec K_F dk_{\|} \frac{c_0}{|\omega|^{\frac{\alpha}{z}}} F_T\left(\frac{c_1 \omega}{k_{\|}^z}, \frac{\omega}{T}\right) \\
& \sim & \int_{FS} |\omega|^{\frac{1 - \alpha}{z}} Y\left(\frac{\omega}{T}\right)
\end{eqnarray}
Here $Y(x)$ is a universal scaling function that is simply related to $F_T$. The integral in the last line is taken over the Fermi surface. (Depending on details there may in addition be a smooth non-singular background).
Note that the $d$-dimensional $\vec K$ integral separates into a $(d-1)$ dimensional integral over the Fermi surface and an one dimensional integral over $k_{\|}$. Performing the latter integral gives the contribution from each `patch' of the Fermi surface to the total singularity in the
tunneling density of states. Thus each patch contributes as a one dimensional system.

If the exponents are independent of the position on the Fermi surface then the Fermi surface integral only contributes a harmless prefactor and a simple power law scaling form results for $N(\omega, T)$. However if the exponents depend on angle then the Fermi surface integral is nontrivial and the result will not be a pristine power law singularity even at frequencies low enough to be in the universal scaling regime. At the lowest frequencies/temperature the Fermi surface integral will de dominated by portions of the Fermi surface with the smallest value of the exponent $\frac{1 - \alpha}{z}$. For instance at $T = 0$ (and specializing to two dimension for simplicity) the integral may be done by saddle point and we find
\begin{equation}
N_s(\omega, T) \sim \frac{|\omega|^{x}}{\sqrt{ln{\frac{1}{|\omega|}}}}
\end{equation}
with $x = min\left(\frac{1 - \alpha}{z}\right)$.

Let us now consider the behavior of the spectral function upon leaving the critical point by tuning an appropriate parameter $g$ away from its critical value $g_c$. As with other critical phenomena we expect that the universal critical singularities will be cut-off at a momentum scale $k_{\|} \sim \frac{1}{\xi}$ and frequency scale $\omega \sim \frac{1}{\xi^z}$. Thus an appropriate scaling ansatz for the spectral function that includes the full crossover out of criticality will take the following form:
\begin{equation}\label{elsfngen}
A(\vec K, \omega, T, g) \sim \frac{c_0}{|\omega|^{\frac{\alpha}{z}}} F\left(\frac{c_1 \omega}{k_{\|}^z}, \frac{\omega}{T}, k_{\|}\xi \right)
\end{equation}
The crossover length scale $\xi$ will diverge as the critical point is approached, presumably as a power law
\begin{equation}
\label{xi}
\xi \sim |g - g_c|^{-\nu}
\end{equation}
However a priori we must again allow for the possibility that $\nu = \nu(\vec K_F)$ is a function of position on the Fermi surface.

Eqn. \ref{elsfngen} is our most general scaling ansatz for the single particle spectral function. As usual this scaling form is applicable for small values of $k_{\|}, \omega, T, |g-g_c|$. It contains a great deal of information about the the universal singularities and crossovers of all single particle properties in the vicinity of the critical point. Let us for instance consider how usual Fermi liquid physics is recovered upon tuning $g$ away from $g_c$ at zero temperature. For concreteness we take this to correspond to $g - g_c > 0$. In this paper we will make the assumption that the low energy physics of the Fermi liquid is part of the universal scaling form for $A(\vec K, \omega)$. If that is not the case then the subsequent scaling analysis of the crossover out of criticality will need to be modified.  With this assumption the scaling form for $A(\vec K, \omega, T = 0, g)$ of Eqn. \ref{elsfngen} at $g > g_c$ must match onto Eqn. \ref{elsfn_lfl} that describes a Landau fermi liquid. This is immediately seen to imply the following singular dependences of the quasiparticle residue $Z$ and the Fermi velocity $v_F$ on $g-g_c$:
\begin{eqnarray}
\label{Zsing}
   Z &\sim & (g- g_c)^{\nu(z - \alpha)} \\
   v_F & \sim & (g-g_c)^{\nu (z - 1)}
   \label{vFsing}
\end{eqnarray}
We emphasize that these equations hold separately for each point on the Fermi surface. If the exponents are angle dependent then the singular structure of $Z$ and $v_F$ will depend on position on the Fermi surface.
The inequality $z \geq \alpha$ guarantees that $Z$ is non-divergent at the critical point as indeed it must be.
The singularity in the Fermi velocity may be interpreted as a singularity in the effective mass of the Landau quasiparticle of the Fermi liquid.
Physically it is reasonable to assume that the effective mass does not vanish as the transition is approached.
For instance at the Mott transition a diverging effective mass is natural and is consistent with the impending localization of the electrons. A vanishing effective mass however is unlikely. This then leads to the inequality
\begin{equation}
z \geq 1
\end{equation}
everywhere on the Fermi surface.

Note that unless $\alpha = 1$ the singularity in $Z$ does not track that in the inverse effective mass $m^* \sim \frac{1}{v_F}$. In general $Z$ is proportional to $1/m^*$ only if the electron self energy is independent of momentum perpendicular to the Fermi surface. Thus if $\alpha \neq 1$ the electron self energy necessarily has nontrivial
singular dependence on $k_{\|}$.

Finally in the Fermi liquid phase the Landau quasiparticle will have a small decay rate $\gamma \propto \omega^2$ on moving away from the Fermi surface. From the scaling form for the electron spectral function we find
\begin{equation}
\label{qprate}
\gamma \sim \xi^z \omega^2
\end{equation}

\section{Scaling hypotheses for thermodynamic and other properties}
\label{othersclhyp}
In this section we will formulate scaling hypotheses for various physical quantities (for instance specific heat) that can often be probed in experiments. Our focus is on the singular contribution to these quantities from the critical Fermi surface modes. In principle there may be singular contributions from other distinct bosonic modes as well (for instance from order parameter fluctuations at a magnetic ordering transition that accompanies the Fermi surface reconstruction). We will make the crucial assumption that the dominant contribution to the singularities in the thermodynamics comes from fluctuations associated with the critical Fermi surface.

\subsection{Thermodynamics}
\label{thermo}
We now consider the singularities in thermodynamic quantities such as the specific heat $C_v$ or spin susceptibility $\chi$
associated with a critical Fermi surface. As usual these can be obtained from a scaling ansatz for the free energy density ${\cal F}(T,B)$ as a function of temperature $T$ and uniform Zeeman magnetic field $B$. We will argue
that there are several qualitatively distinct scaling models for the thermodynamics.

In thinking about
the scaling of the free energy it is extremely instructive to consider the fate of thermodynamic quantities as the critical point is approached from the Fermi liquid side. The singularity in the effective mass described by Eqn. \ref{vFsing} determines the singularity in the coefficient $\gamma = \frac{C_v}{T}$. Specifically we have for $g > g_c$ and $T \rightarrow 0$
\begin{equation}
C_v \sim T \int_{FS} \frac{1}{v_F}
\end{equation}
so that
\begin{equation}
\label{gammadiv}
\gamma  \sim \int_{FS} (g-g_c)^{-\nu (z - 1)}
\end{equation}
Note that unless the exponents are angle independent this is again not a pristine power law divergence even in the universal scaling regime. Rewriting this as
$\gamma \sim \int_{FS} \xi^{z-1}$ suggests a natural guess for the singularity in the specific heat right at the critical point. We simply replace $\xi$ by $T^{-{\frac{1}{z}}}$ to get
\begin{equation}
\label{Csing}
C_v \sim \int_{FS} T^{\frac{1}{z}}
\end{equation}
More generally the full singularity of the (zero field) specific heat for small $T$ and $|g - g_c|$ is expected to be described by the scaling form
\begin{equation}
\label{Cscal}
C_v(T, g) \sim \int_{FS} T^{\frac{1}{z}}{\cal C}\left(T\xi^{z}\right)
\end{equation}
where ${\cal C}$ is a universal scaling function satisfying ${\cal C}(x \rightarrow 0) \sim x^{\frac{z- 1}{z}}$
and ${\cal C}(x \rightarrow \infty) \sim 1$. In the special case in which $z = 1$ everywhere on the critical Fermi surface there is no singularity in the specific heat at the transition. Note also that in general with angle dependent exponents scaling functions like ${\cal C}$ may also themselves depend on position on the Fermi surface.

As with the tunneling density of states discussed above, if the exponents are angle dependent then the asymptotic singular behavior is dominated by some portions of the Fermi surface. For the specific heat it will be dominated by the region with largest $z$. For instance (in two dimensions) we find right at $g_c$
\begin{equation}
C_v \sim \frac{T^{\frac{1}{z_{max}}}}{\sqrt{ln\frac{1}{T}}}
\end{equation}
where $z_{max}$ is the maximum value of $z$ on the Fermi surface.
In general the singularities in different physical quantities will be dominated by different portions of the Fermi surface as they will involve different combinations of exponents.

The scaling ansatz of the specific heat immediately leads to a scaling form for the entropy $S$  near the quantum critical point:
\begin{equation}
S \sim \int_{FS} T^{\frac{1}{z}}{\cal S}\left(T\xi^{z}\right)
\end{equation}
A useful experimental probe is the Gruneisen parameter $\Gamma$ defined as the ratio of the molar specific heat at constant pressure $ c_p$  and the volume thermal expansion $\beta = \frac{1}{V}\frac{\partial V}{\partial T}$:
\begin{equation}
\Gamma = \frac{\beta}{c_p}
\end{equation}
The singularities of $\Gamma$ at various quantum critical points were examined theoretically in Ref. \onlinecite{garst}. The volume thermal expansion can be related to the derivative of the entropy with respect to pressure. Assuming that the pressure changes couple linearly to the tuning parameter $g-g_c$, the scaling form of the entropy determines the singularity in $\beta$. In the presence of a critical Fermi surface we find
\begin{equation}
\beta \sim \int_{FS} T^{\frac{1-\frac{1}{\nu}}{z}}
\end{equation}
Again if the exponents were angle dependent this would be dominated by some portions of the Fermi surface which could in principle be different from the portions that dominate the specific heat. Thus with angle
dependent exponents the Gruneisen parameter would have a complicated temperature dependence.  If the exponents are angle independent however we find $\Gamma \sim T^{-\frac{1}{z\nu}}$.

A heuristic way to understand the scaling ansatz for the specific heat or entropy is to recognize that each local portion of the critical fermi surface contributes as a one dimensional system. For a critical point in one space dimension the specific heat singularity is $C_v \sim T^{\frac{1}{z}}$. The full specific heat is an integral over these contributions from each such portion.

Let us now consider the scaling of the susceptibility. For $g > g_c$ in the Fermi liquid the susceptibility is not determined just by the quasiparticle density of states but is corrected by a Landau interaction parameter and may be written in the form
\begin{equation}
\chi \sim \frac{\rho_o}{1 + F^a}
\end{equation}
Here $\rho_0 \sim \int_{FS} \frac{1}{v_F}$ is the quasiparticle density of states at the Fermi surface and $F^a$ is a Landau parameter. We can now distinguish four qualitatively distinct situations which we will denote Models I, II, III, and IV respectively. In Model I  $F^a$ remains finite as the quantum critical point is approached. In this case the susceptibility will diverge at the quantum critical point following the divergent density of states.
Such a possibility was proposed by Brinkman and Rice in their theory of the Mott transition\cite{BrRice}.
In Model II $F^a$ also diverges on approaching the quantum critical point in exactly the same way as the density of states so that $\chi$ stays non-singular. As described elsewhere\cite{tsmott} this is precisely realized in slave particle gauge theoretic descriptions of such quantum phase transitions (for instance in the theory of Ref. \onlinecite{svsrev} for a Fermi volume changing transition in the Kondo lattice). In Model III, $F^a$ diverges more strongly than the density of states so that the susceptibility goes to zero at the transition. This might be expected, for instance, at the Mott transition if the insulating side has a full spin gap. Finally in Model IV
$F^a$ approaches $-1$ on approaching the quantum critical point. This corresponds to a ferromagnetic instability. Such ferromagnetic transitions will not be considered in this paper (nor will the analogous Pomeranchuk instabilities to various other ordered states - see Refs. \onlinecite{nematic} for a recent discussion). They are likely to have rather different structure from the critical points discussed in this paper.

A scaling ansatz for the susceptibility which includes all the crossovers near the critical point can be readily written down for Model I. Following the same logic as above for the specific heat we write
\begin{equation}
\label{chiscal}
\chi(T, g) \sim \int_{FS} T^{\frac{1}{z}-1}X\left(T\xi^{z}\right)
\end{equation}
where $X$ is a universal scaling function satisfying $X(x \rightarrow 0) \sim x^{\frac{z- 1}{z}}$
and $X(x \rightarrow \infty) \sim 1$. Again with angle dependent exponents the scaling function $X$ may itself also have angle dependence. At the lowest temperatures $\chi$ will again be dominated by the same portion of the Fermi surface that dominates the specific heat. Thus we find
\begin{equation}
\chi \sim \frac{T^{\frac{1}{z_{max}}- 1}}{\sqrt{ln\frac{1}{T}}}
\end{equation}

In general we note that the Wilson ratio $\frac{\chi T}{C}$ will have a complicated temperature dependence if the exponents and scaling functions have angle dependence. However at the lowest temperatures it will go to a constant
as both $\chi$ and $C_v$ are dominated by the same portions of the Fermi surface.

\subsection{Two particle correlators and transport: The critical $2K_f$ surface}
\label{2kf}
Clearly the scaling hypotheses can be generalized to various other physical quantities. Of particular interest are
two particle properties such as the spin or charge density correlators, and transport. Consider for instance the dynamical spin susceptibility $\chi''(\vec K, \omega)$. With a sharp Fermi surface it is natural to expect that this will be gapless at various momenta $\vec K$ in the Brillouin zone. In particular it may be expected to have sharp structure at various ``$2\vec K_f$" wavevectors which connect parallel portions of the Fermi surface. We expect that right at any given $\vec Q$ wavevector that connects such parallel portions the dynamical spin susceptibility at $T = 0$ will have scale invariant singular dependence on frequency:
\begin{equation}
\chi"(\vec Q, \omega) \sim |\omega|^{y}
\end{equation}
with $y$ in principle different for different $\vec Q$ momenta. The locus of such preferred momenta
$\vec Q$ will define a surface (or surfaces) in momentum space that is distinct from the Fermi surface. We will refer to this as the $2K_f$ surface. Unlike at bosonic quantum critical points the spin correlations are critical along this entire surface in momentum space.

For small deviations of the momentum from the $2K_f$ surface we expect a singular dependence on the susceptibility on the momentum deviation. This will be captured by a scaling form
\begin{equation}
\chi"(\vec k, \omega, T) \sim \omega^y f\left(\frac{a\omega}{q_{\|}^{z_2}}, \frac{\omega}{T} \right)
\end{equation}
where $q_{\|}$ is the deviation of the momentum $\vec k$ from the $2K_f$ surface at the point of closest approach, and $f$ is a universal scaling function. ($a$ is a non-universal constant). For generality we have included a nonzero temperature. We have also introduced a new dynamical critical exponent $z_2$. Again a priori we should allow for $z_2$ to vary as we move around the $2K_f$ surface. It is not a priori clear  how $z_2$
is related to $z$.

Next we briefly consider the much more delicate issue of the criticality of transport properties. Quite generally we expect  that the electrical conductivity $\sigma(\omega, T)$ at the critical point will satisfy
\begin{equation}
\sigma(\omega, T) \sim \int_{FS} T^{-\mu} \Sigma\left(\frac{\omega}{T}\right)
\end{equation}
with $\mu$ a universal exponent and $\Sigma$ a universal scaling function. We point out that in the presence of a critical Fermi surface there is no reason to expect that the conductivity exponent $\mu$ will be the same as at bosonic quantum critical points. Thus in $d = 2$ there is no serious reason to assume that the conductivity at such fermionic quantum critical points will be a universal constant. See Section \ref{cuprates} for a plausible argument for $\mu = 1$ in the cuprate materials in the strange metal region.

\section{Consequences of angle dependent exponents: Metals with ``Fermi arcs"}
\label{fermiarcs}

We now argue that angle dependent exponents lead to some remarkable phenomena in the vicinity of the quantum critical point. In particular they very naturally lead to metals with ``gapless Fermi arcs" at finite temperature.
In general angle dependence of exponents implies that on leaving the critical point different portions of the Fermi surface will emerge out of criticality at different length/energy scales. Thus at a nonzero temperature slightly away from the critical point, some portions of the Fermi surface will have emerged out of the `quantum critical' regime while others will not have. This leads to finite temperature crossovers that are much richer than near other familiar quantum critical points.

For
concreteness we consider the Mott transition, say as a function of interaction strength at half-filling on a frustrated lattice. In the Mott insulator there will a gap to single particle excitations at zero temperature. As argued in the Introduction this gap is expected to close on an entire surface in momentum space on approaching the critical point.
This surface will match on to the Fermi surface of the metal on the other side of the transition. From the scaling ansatz
of Eqn. \ref{elsfngen} it follows that the gap will vanish as
\begin{equation}
\label{gapscal}
\Delta \sim |g - g_c|^{z\nu}
\end{equation}
With angle dependent exponents the gap will vanish with different power laws at different portions of the Fermi surface. Now consider the finite temperature crossovers in the vicinity of the quantum critical point. Right at $g = g_c$ the finite temperature physics is described by the universal scaling forms of previous sections. The resulting state may be described as a `strange metal' with a critical Fermi surface. Upon moving into the Mott insulator by decreasing $g$ from $g_c$, a portion of the Fermi surface with gap $\Delta$ will emerge out of the quantum critical region at a temperature $T(\hat{K_F}) \sim \Delta(\hat{K_F})\sim |g - g_c|^{z\nu}$. With angle dependent exponents this is a different crossover temperature for different portions of the Fermi surface. Thus for any fixed $g$ different from $g_c$ there will be an extended crossover regime as a function of temperature where the system is emerging out of the quantum critical region. This regime will occur for $T_{
 min} < T < T_{max}$ where
$T_{min} \sim |g-g_c|^{(z\nu)_{max}}$ and $T_{max} \sim |g - g_c|^{(z\nu)_{min}}$. In this intermediate temperature regime the Fermi surface is partially gapped. The properties will be that of a metal with temperature dependent Fermi arcs. The arc length will shrink with decreasing temperature.

On the other side of the transition with decreasing temperature coherent quasiparticles will emerge out of the critical Fermi surface. Again the coherence will first be established in portions of the Fermi surface with the smallest $z\nu$ and will gradually spread to other portions. Clearly the portions of the Fermi surface which first get gapped in the Mott side are also the portions where coherent quasiparticles first emerge in the Fermi liquid side.

\begin{figure}
\includegraphics[width=8cm]{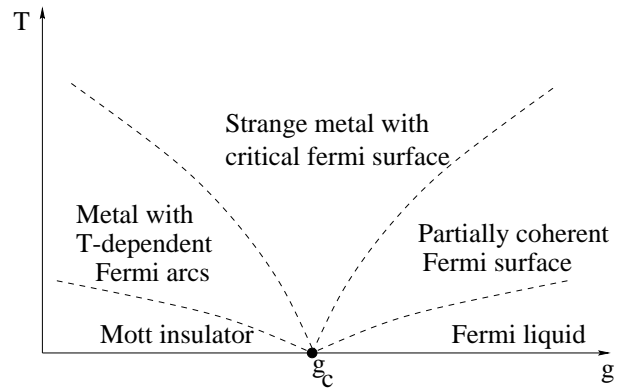}
\caption{Finite temperature crossovers near a second order Mott transition with angle dependent exponents on the critical Fermi surface.} \label{mttxovr}
\end{figure}

A schematic finite temperature crossover diagram is shown in Fig. \ref{mttxovr}. We note a remarkable similarity with the phenomenology of the cuprate metals near optimal doping.

\section{Application to cuprate metals}
\label{cuprates}
Motivated by the discussion in the previous section we now provide a tentative application of the scaling ideas of this paper to the cuprate metals. Consider the ``underlying normal" ground state of the cuprates as the doping $x$ changes from underdoped to overdoped. What might one mean by the ``underlying normal" ground state? Theoretically given some model Hamiltonian $H$ that describes the cuprates we can take the lowest energy non-superconducting ground state and formally imagine tuning away all instabilities to superconductivity. It is tempting to associate this with the state accessed in experiments by suppressing the superconductivity in a magnetic field. However some caution is needed as the magnetic field might also cause a phase transition in the ``normal" ground state.

In the overdoped side it seems very likely that the underlying normal state is a Landau Fermi liquid with a large Fermi surface satisfying Luttinger's theorem\cite{overdpfs}. On the other hand the normal ground state at underdoping seems very unlikely to be smoothly connected to such a large Fermi surface Fermi liquid. Recent experiments on underdoped cuprates at low $T$ and high magnetic fields find a metallic state showing quantum oscillations\cite{underdposc}. Thus the underdoped normal ground state apparently has a sharp Fermi surface of fermionic charge carriers. The frequency and other features of the quantum oscillations suggest that this Fermi surface is rather different from that in the overdoped side. Several possible candidate states have been discussed theoretically for the underdoped normal ground state which all have small hole pockets along the four diagonal directions of the Brillouin zone of the square lattice. Examples include
\begin{enumerate}
\item
Metal with commensurate $(\pi, \pi)$ antiferromagnetic (AF) order with 2 small hole pockets\cite{zhang} centered at $(\pm \pi/2, \pm \pi/2)$

\item
Staggered flux (also known as $d$-density wave) state which also has 2 small hole pockets\cite{s-flux} centered at
$(\pm \pi/2, \pm \pi/2)$

\item
Metal with incommensurate spin density or charge density order\cite{kivelson,hudson}.

\item
Fractionalized Fermi liquid states\footnote{Such states do not satisfy the conventional Luttinger theorem and hence cannot be Landau Fermi liquids; they necessarily need to also have topological or other such exotic order and associated fractionalized excitations. See Refs. \onlinecite{oshi,flstar}} with four small hole pockets\cite{rice,kotliar} and which preserve all microscopic symmetries.

\item
Other even more exotic states with fractionalized excitations such as the holon-hole metal of Ref. \onlinecite{acl}.
\end{enumerate}
The first two example break symmetry leading to doubling of the unit cell. The third also breaks lattice translation symmetry while the last two are exotic states that violate the standard Luttinger theorem.

It is at present not clear which (if any) of these possible states is realized in the cuprates. In this paper we will not attempt to resolve this issue. We will take the point of view that different theoretical models of doped Mott insulators might realize these different possibilities. Whichever one is realized the issue of how it evolves into the overdoped large Fermi surface Fermi liquid needs to be confronted.

We will make the fundamental assumption that as $x$ is decreased from the overdoped side there is a critical $x_c$ at which the large Fermi surface disappears. Upon further decreasing $x$, the underdoped metallic state state with its fundamentally different Fermi surface appears. We will further assume that this  quantum phase transition is second order. We may then use the scaling theory developed in this paper to describe various finite temperature phenomena in the vicinity of $x_c$. We will see that this provides a very interesting conceptual framework that naturally encapsulates several of the mysterious normal state phenomena in the cuprates.

The assumption that the large Fermi surface disappears through a second order quantum phase transition leads to the existence of a critical large Fermi surface at $x = x_c$ but with no sharp Landau quasiparticle. For concreteness it will be useful to consider situations in which the underdoped metal for $x < x_c$ is a broken symmetry state with a doubled unit cell and 2 small hole pockets - for instance a state with staggered flux/ddw order. (However we expect the general framework sketched below to be easily generalizable if a different broken symmetry or a more exotic state is realized). On approaching $x_c$ from the underdoped side then, the small hole pockets must disappear together with the broken symmetry leading to the doubling of the unit cell. We are then naturally lead to expect that the small hole pockets remain sharply defined at $x_c$ but again with no sharp quasiparticle peak as in the other cases discussed in previous sections.

\begin{figure}
\includegraphics[width=8cm]{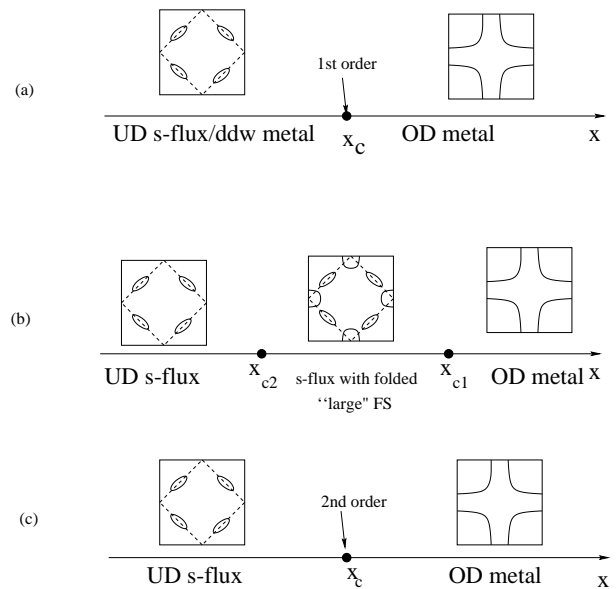}
\caption{Various possible evolutions between a small Fermi surface underdoped metal (such as the staggered flux/ddw state) and a large Fermi surface overdoped Fermi liquid. In (a) there is a direct first order transition. In (b) the transition proceeds through an intermediate phase which has staggered flux order but has a folded version of the large Fermi surface. This state has both hole and electron pockets. The transition at $x_{c2}$ is a Lifshitz transition while the one at $x_{c1}$ involves the development of the broken symmetry but no extra reconstruction of the large Fermi surface. This transition will be described by a theory along the lines of Ref. \onlinecite{Hertz} and will only have weak departures from Fermi liquid theory. In (c) there is a direct second order transition where the Fermi surface reconstruction from large to small accompanies the development of the broken symmetry. The critical point at $x_c$ will be strongly non-fermi liquid like.} \label{hitcpdia2}
\end{figure}

In passing we note that the disappearance of the large Fermi surface is not simply due to the broken symmetry assumed in the underdoped side. If that were the case the resulting underdoped fermi surface will simply be a folded version of the overdoped one which will look very different from the assumed one. This is elaborated in Fig. \ref{hitcpdia2}, and discussed further in Section \ref{disc}. Our assumption is that Fig. \ref{hitcpdia2}c with a direct second order transition where the Fermi surface abruptly changes from ``large" to ``small" is realized. (In contrast Fig. \ref{hitcpdia2}b is what would be expected if the transition out of the overdoped Fermi liquid only involves the onset of the broken symmetry order parameter.) Indeed we may imagine that the ``primary" transition is associated with the death of the large Fermi surface, and that the broken symmetry for $x < x_c$ is a low energy instability of the state that results once the large Fermi surface disappears.

The critical point at $x_c$ has sharply defined Fermi surfaces corresponding both to the overdoped and underdoped Fermi surfaces so that the electron spectral function will have universal singularities at both Fermi surfaces. We may then use the scaling hypotheses discussed in previous sections to discuss the single particle spectrum in the universal scaling regime in the vicinity of $x_c$. In what follows we will use a subscript $l$ to denote exponent functions associated with the large Fermi surface(LFS) and a subscript $s$ for the small Fermi surface(SFS).

Consider first the singularities associated with the large Fermi surface. As discussed in previous sections at non-zero $T$ , this will lead to a quantum critical strange metal  at $x = x_c$, to a metal with Fermi arcs along portions of the large Fermi surface for $x < x_c$, and to a partially critical large Fermi surface at $x > x_c$.
All of this is strikingly reminiscent of what is known about the cuprates from ARPES experiments\cite{hitcarpes}.

Our discussion also implies that singularities corresponding to the small hole pockets. However there is no sign of these small hole pockets in the ARPES experiments. This can be accounted for by assuming that the critical exponent $\alpha_s$ (see Eqn. \ref{elsfn_cfs}) is negative everywhere on the small Fermi surface associated with the hole pocket. Then the critical spectral function ({\em i.e} at $x = x_c$) right at the small Fermi surface has the structure
\begin{equation}
A(\vec K \in SFS, \omega) \sim |\omega|^{\theta_s}
\end{equation}
with $\theta_s = |\alpha_s|/z_s > 0$ . This is a vanishing cusp singularity at the Fermi energy, and hence may easily evade detection in ARPES experiments. If on the other hand $\alpha > 0$ on the large Fermi surface there will be a
diverging power law singularity in $A(\vec K, \omega)$ which will be much more readily visible in photoemission experiments.

\begin{figure}
\includegraphics[width=8cm]{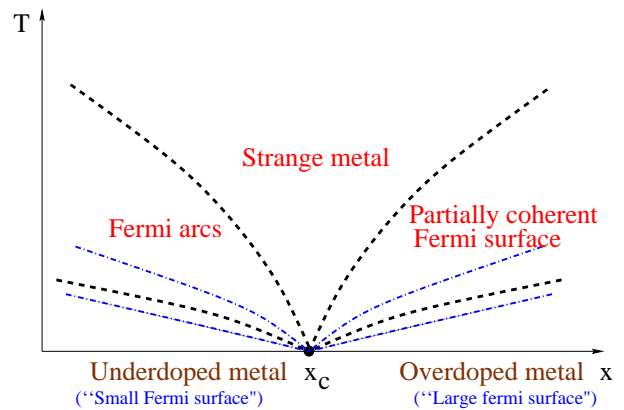}
\caption{Finite temperature crossovers near the proposed quantum critical point between the
 overdoped ``large Fermi surface" metal and an underdoped ``small Fermi surface" metal. The black dashed lines mark the crossovers associated with the large Fermi surface and the blue dash-dot lines those of the small Fermi surface. It is suggested that the critical singularities in the spectral function associated with the small Fermi surface are too weak to be resolved in photoemission experiments.} \label{hitcxovr}
\end{figure}

A schematic crossover phase diagram is shown in Fig \ref{hitcxovr}. For $x < x_c$ the large Fermi surface will begin its crossover out of the quantum critical strange metal at a temperature $T^* \sim |x- x_c|^{(z_l\nu_l )_{min}}$. Below this temperature the large Fermi surface will only consist of gapless Fermi arcs. At a lower temperature scale $T^{**} \sim |x- x_c|^{(z_l\nu_l)_{max}}$ the destruction of the large Fermi surface is complete. The small Fermi surface will begin its crossover out of the strange metal region at a temperature $\sim |x- x_c|^{(z_s\nu_s )_{min}}$. So long as
$(z_s \nu_s)_{min} > (z_l \nu_l)_{min}$, the $T^*$ line will only involve forming a pseudogap along portions of the large Fermi surface. We may therefore identify this with the experimentally determined pseudogap line.

At the lowest temperatures the Fermi surface will consist entirely of small hole pockets. However with $\alpha_s$ negative the quasiparticle residue $Z \sim |x - x_c|^{\nu_s(z_s - \alpha_s)}$ will grow slowly with decreasing doping from $x_c$. In many models of the underdoped metal, the residue $Z$ also goes to zero as $x$ goes to zero. Thus $Z$ may never become very big on the small Fermi surface. This will affect the visibility of this Fermi surface in ARPES but not in quantum oscillations or in thermodynamics.

We emphasize that much of the physics in the strange metal region and the initial crossover at $T^*$ are properties of the critical large Fermi surface. The details of the underdoped metal - in particular exactly what symmetry is broken if any, precise structure of the small Fermi surface, and so on - are suggested to be determined at low temperatures (below
$\sim |x- x_c|^{(z_s\nu_s )_{min}}$) much smaller than the $T^*$ temperature at which the large Fermi surface starts
its crossover out of the quantum critical strange metal region. This is the meaning of the statement that the primary transition is that associated with the death of the large Fermi surface.

Before concluding this section a few comments are in order. First in the cuprates there is no sign of the quasiparticle mass diverging as optimal doping is approached from either side. This implies that $z = 1$ everywhere on the critical Fermi surfaces. The angle dependence of the gap must then be due to the exponent $\nu$. Finally we consider transport properties.
Consider the $T$-dependence of the resistivity in the overdoped Fermi liquid phase.
We know (see Eqn. \ref{qprate}) that the quasiparticle scattering rate at a non-zero $T$ is $\gamma \sim \xi^z T^2$ with $z = 1$. In the Fermi liquid phase if we simply take this to also be the transport scattering rate $\gamma_{tr}$ then the temperature dependence of the resistivity $\rho(T)$  will behave as
\begin{equation}
\rho(T) \sim A T^2
\end{equation}
with
\begin{equation}
A \sim \xi
\end{equation}
Note that with $z = 1$ scaling the density of states is not critically enhanced so that
the scattering rate determines the scaling of the $A$ coefficient.
At criticality $\xi$ will diverge, and a natural guess for the $T$-dependence is given by replacing $\xi$ by $T^{-\frac{1}{z}} \sim \frac{1}{T}$. This then gives
\begin{equation}
\rho(T) \sim T
\end{equation}
at the critical point. Such a linear resistivity is of course well known in the strange metal region in the cuprates.

Of course at this stage these ideas are clearly to be regarded as tentative. However the line of thinking about the cuprates suggested in this section is rather appealing. It describes a strange metal regime at finite $T$ around optimal doping which crosses over into a Fermi arc metal in the underdoped region. Furthermore the arc length shrinks as $T$ is reduced. It also allows for recovering a conventional Landau Fermi liquid at overdoping. Finally it allows for a reconciliation between Fermi arcs in ARPES at intermediate $T$ and quantum oscillations at low $T$ in the underdoped side. We hope that it provides a framework to usefully address normal state phenomena in the cuprates.

\section{Slave particle calculation with a critical Fermi surface}
\label{slave}
What kinds of calculations might one do to discuss the kind of quantum criticality considered in this paper? Currently there exists one technique which is based on a slave particle description of
phase transitions associated with the disappearance of a fermi surface. For the Kondo lattice such a description was introduced in Ref. \onlinecite{svsrev} and is known as the Kondo breakdown model (for some representative subsequent work see Refs. \onlinecite{kb}). A similar slave particle theory can also be constructed for the Mott transition on a non-bipartite lattice using a `slave rotor' formulation\cite{florens} and including fluctuations. In this Section we first show that a simple slave particle calculation provides theoretical evidence for the notion of a critical Fermi surface.  We will also put such slave particle theories in the context of the general scaling ideas discussed in this paper.

Consider a one band Hubbard model at half-filling on a non-bipartite lattice such as the triangular lattice:
\begin{equation}
H = -t\sum_{<rr'>} \left(c^\dagger_{r}c_{r'} + h.c \right) + U\sum_r \left(n_r - 1\right)^2
\end{equation}
where $c_r$ destroys a spinful electron at site $r$ of a triangular lattice. $n_r = c^\dagger_rc_r$ is the electron number at site $r$. $U > 0$ is an on-cite repulsion. For large $g = t/U$ the ground state is a stable Fermi liquid metal. For small $t/U$ a Mott insulator results. Clearly there needs to be a Mott metal insulator transition at some critical value of $g_c = \left(t/U\right)_c$. Deep in the insulating phase the low energy physics is described by the nearest neighbor Heisenberg model which has $120^0$ coplanar Neel order.  However it is hardly clear that this Neel order survives on approaching close to the Mott transition in the insulating phase. Indeed motivated by recent experiments on the organic material $\kappa-(ET)_2Cu_2(CN)_3$ (believed to be described by a one band Hubbard model on an isotropic triangular lattice), it has been suggested\cite{lesik,leesq} that a spin liquid insulating state with a fermi surface of neutral fermionic spinons is realized in the immediate vicinity of the Mott transition. Such a spin liquid Mott insulating state allows for a direct second order Mott transition into the metallic Fermi liquid state that obtains at smaller $U/t$. Here we briefly consider a simple mean field theory of this transition and show that a critical Fermi surface is indeed realized at the transition.  More sophisticated calculations and specific predictions for $\kappa-(ET)_2Cu_2(CN)_3$ will be discussed in a companion paper\cite{tsmott}.

The Mott transition and the spin liquid phase are conveniently discussed using the slave rotor representation of Ref. \onlinecite{florens}. We write
\begin{equation}
c_{r\alpha} = e^{i\phi_r} f_{r\alpha}
\end{equation}
with $e^{i\phi_r} \equiv b_r$ a spin-$0$ charge-$e$ boson, and $f_{r\alpha}$ a spin-$1/2$ charge-$0$ fermionic spinon. We start with a mean field description in which the spinons are non-interacting and form a Fermi surface. If the boson $b_r$ is condensed ($<b_r> \neq 0$) the result is the Fermi liquid phase of the electrons. If the boson is gapped (and hence uncondensed) a spin liquid Mott insulator with a spinon Fermi surface results. The phase transition at $g_c$ between the two phases is driven by the condensation of the boson $b_r$. A low energy effective theory for the transition is given by the action
\begin{eqnarray}
S & = & S_b + S_f + S_a + S_{bf}\\
S_b & = & \int d^2x d\tau |\left(\partial_\mu - i a_\mu \right) b|^2 + V\left(|b|^2\right) \\
S_f & = & \int_{\vec x, \tau} \bar{f}_\alpha \left(\partial_{\tau} - \mu_f + ia_0 - \frac{(\vec \nabla + i \vec a)^2}{2m_f}\right)f_\alpha \\
S_a & = & \int_{\vec x, \tau}\frac{1}{e_0^2}\left(\epsilon_{\mu\nu\lambda \partial_\nu a_\lambda} \right)^2
\end{eqnarray}
The $a_{\mu}$ is a $U(1)$ gauge field that appears due to the redundancy introduced by the slave rotor representation of the electron operators. The potential $V\left(|b|^2\right)$ may simply be taken to be of the usual form $r|b|^2 + u|b|^4$. The last term $S_{bf}$ represents residual short range interactions between the bosons and fermions. The most important of these is a coupling between $|b|^2$ and a suitable fermion bilinear.

In this paper we consider a simple `mean field' approximation in which we ignore the gauge fields but not other interactions. Fluctuation effects are examined in Ref. \onlinecite{tsmott}. The boson condensation transition of action $S_b$ is then in the $3D$ XY universality class. The interaction terms in $S_{bf}$ can be shown to be irrelevant at the corresponding fixed point\cite{tsmott}. Thus the bosons and spinons are decoupled in the absence of gauge interactions. The electron Greens function is then readily calculated by convolving the spinon and boson Greens function. In the Fermi liquid side a quasiparticle pole is obtained with a residue
$Z \sim |<b>|^2$. Clearly $Z$ vanishes on approaching the Mott transition due to the vanishing condensate fraction
as $Z \sim |g - g_c|^{2\beta}$ where $\beta$ is the order parameter exponent for the $3D$ XY model.
In this approximation the quasiparticle effective mass does not diverge and stays finite as the transition is approached. (This will however change once gauge fluctuations are included\cite{tsmott}). At this mean field level the electron spectral function at the critical point is readily calculated (see Appendix), and we find
\begin{equation}
A_c^{(mf)}(\vec K, \omega) \sim |\omega|^{\eta} F^{(mf)}\left(\frac{c_0 \omega}{k_\parallel}\right)
\end{equation}
Thus in this simple mean field calculation there is indeed a critical Fermi surface at which the electron spectral function has singularities. Further it satisfies scaling with the exponents $z = 1$ and $\alpha = -\eta$ (with $\eta$ the anomalous exponent of the boson field at the $3D$ XY fixed point). Note that these exponent values are consistent with Eqns. \ref{Zsing} if we use the known exponent equality $\beta = \frac{\nu(1 + \eta)}{2}$ of the $3d$ XY model.

This simple mean field result will be modified upon including gauge fluctuation effects\cite{tsmott}.  Here we merely note that the scaling form for the critical single particle spectral function above continues to hold (but with extra logarithms). Further the fluctuations lead to a divergence of the quasiparticle effective mass on approaching the transition from the Fermi liquid side but the spin susceptibility stays constant. Thus this transition is an example of a Model II transition (in the nomenclature of Section \ref{thermo}). However in this example the universal scaling function for the crossover out of criticality does not in fact contain the low energy physics of the Fermi liquid\cite{tsmott}. This may be a general limitation of slave particle gauge theories.

A key point to notice is that the critical exponents are {\em angle independent} at this particular Mott transition. This may be a further limitation of such simple slave particle descriptions, and is perhaps why phenomena such as fermi arc metals have not been easily found thus far within the slave particle framework. The possibility of angle dependent exponents is further discussed below.

\section{Discussion}
\label{disc}
In this concluding section we discuss several issues brought out by this paper, and the implications of these ideas for experiments on various quantum critical systems.

\begin{enumerate}
\item
{\em Can the phase transtions considered at all be second order?}

The most fundamental assumption we have made is that the phase transitions we have discussed can be second order. Can a Fermi surface disappear through a continuous second order phase transition? Theoretically there are concrete examples of such second order phase transitions, such as in the Kondo breakdown model of Ref. \onlinecite{svsrev} or at the Mott transition discussed in Section \ref{slave}. However in both examples when the electron Fermi surface disappears the state that results is rather exotic and has a Fermi surface of neutral spin-$1/2$ fermionic spinons. It is not known theoretically whether in more general cases a direct second order transition can occur. For instance in the cuprate context the natural possibilities in Fig. \ref{hitcpdia2} might be expected to be (a) or (b). Fig \ref{hitcpdia2}c requires that the Fermi surface reconstruction from ``large" to ``small" occur at the same point as the onset of the assumed broken symmetry of the underdoped side. So a priori it is not clear that it can be second order. Similar concerns can be raised, for instance for the Mott transition between a paramagnetic Fermi liquid and an antiferromagnetic Mott insulator.

There are two reasons why, despite this natural expectation, such a second order transition might actually be possible. First empirically in heavy electron critical points it appears that something similar may in fact be going on: a drastic Fermi surface reconstruction quite likely accompanies onset of a broken symmetry. Second, theoretically in much simpler problems in quantum magnetism, it has been possible to demonstrate that second order transitions can exist between two different states with very different kinds of order contrary to natural expectations\cite{deccp}. Thus it is also a priori not clear that phase diagrams like Fig. \ref{hitcpdia2}c are not possible.

In passing we note (based on intuition from the theory of deconfined criticality) that even if Fig. \ref{hitcpdia2}c were possible it may easily change to Fig. \ref{hitcpdia2}b in the presence of a  ``relevant" perturbation. For instance an external orbital magnetic field may split the single phase transition of \ref{hitcpdia2}c into the two phase transitions of \ref{hitcpdia2}b thereby opening up an intermediate field induced phase with a folded Fermi surface and electron pockets. Such a possibility might influence details of experiments done at high magnetic fields (such as quantum oscillations).

\item
{\em Can the scaling exponents have angle dependence on the critical Fermi surface?}

As we argued angle dependence of the scaling exponents lead to a number of interesting new phenomena (such as metals with Fermi arcs). The model for the Mott transition to a gapless spin liquid insulator in Section \ref{slave} did not have angle dependence of the exponents. Some intuition can be obtained from models of bosons that have critical power law correlations at surfaces in momentum space (``Bose surfaces"). Two such models have been studied recently\cite{dbl,bll}. In both cases angle dependent exponents are found (rather generically) on a lattice. This suggests that angle dependence of exponents might well happen in electronic systems with a critical Fermi surface.

More generally we anticipate that a variety of universality classes are possible - some with angle dependence in exponents and some without. In the former case the exponent functions presumably also depend on the shape of the Fermi surface, and hence may deform continuously with changes in the Fermi surface shape.

\item
{\em Implications for experiments}

The best studied examples of quantum critical points are in the heavy fermion systems where there is some evidence for a strong Fermi surface reconstruction accompanying the magnetic ordering transition.
Unfortunately it has apparently not been possible to directly obtain information on the single particle spectral function at the critical point.  In the future it will be interesting to look for a sharply defined critical Fermi surface with a scale invariant spectral function through photoemission as a direct test.

At heavy electron critical points there certainly will also be fluctuations associated with the magnetic order parameter. These may be expected to contribute to singularities in thermodynamics as with other critical bosonic modes. In our discussion of the thermodynamics we have focused on the contribution from the critical Fermi surface. How these interact with the critical bosonic order parameter fluctuations is largely an open question. We may cautiously expect that (as in the Fermi liquid phase itself) the large number of degrees of freedom associated with the critical Fermi surface make their contribution dominant over the order parameter fluctuations.

As discussed in Section \ref{2kf}  the critical Fermi surface itself will lead to spin correlations that are critical at a sharp $2K_f$ surface. This means that critical spin scattering will be seen not just at isolated points in momentum space but along lines or surfaces depending on spatial dimension. In $CeCu_{6-x}Au_x$ tuned to its magnetic critical point, there is evidence for critical spin fluctuations not just at the ordering wavevector but along entire lines in momentum space. We suggest that this may be understood as a $2K_f$ surface related to an underlying critical Fermi surface.

The scaling of the specific heat provides some information on various critical exponents. In $YbRh_2\left(Si_{1-x}Ge_x\right)_2$ when tuned by a magnetic field $B$ to the quantum critical point the specific heat follows the scaling form at very low temperatures\cite{hfcscal}
\begin{equation}
\label{yrscscal}
\frac{C_v}{T} \sim \frac{1}{b^{\frac{1}{3}}}\Phi\left(\frac{T}{b}\right)
\end{equation}
with $b = B - B_c$ where $B_c$ is the critical magnetic field. This may be compared with the scaling form in Eqn. \ref{Cscal} expected with a critical Fermi surface
If we tentatively assume that the exponents are angle independent then we get $z = 3/2$, $\nu = 2/3$. We emphasize that the scaling in Eqn. \ref{Cscal} is distinct from that at bosonic critical points, and the $z$ we extract has a rather different meaning. The Gruneisen parameter was also measured\cite{grun} and fit to a diverging power law. The exponent is however apparently inconsistent with the scaling of Eqn. \ref{yrscscal} found in an earlier experiment\cite{hfcscal}. The origin of this discrepancy is presently unclear. More experimental studies are needed to clarify this issue. Future direct measurements of the electron spectral function can in principle check for the validity of these scaling exponents asymptotically close to the Fermi surface.

\end{enumerate}

In closing we reiterate some of our main points. We have argued that at a quantum critical point where a Fermi surface disappears the Fermi surface will continue to be sharply defined even though there is no Landau quasiparticle. The presence of such a  critical Fermi surface will alter the structure of the scaling phenomena expected near the critical point. We have formulated scaling hypotheses for a variety of different physical quantities which can in principle be checked in experiments. Several differences with bosonic quantum critical points were emphasized. We point out that unusual phenomena such as metals with temperature dependent Fermi arcs can occur in a natural way near such a quantum critical point. We discussed normal state phenomena in the cuprates within our scaling framework and the assumption that the
large Fermi surface of the overdoped Fermi liquid disappears below some critical doping through a continuous transition to an underdoped metal. Phenomena near heavy electron critical points were also examined. For the future several important theoretical questions of course remain to be addressed about such critical fermi surfaces. We hope that the scaling theory we have developed will set the stage for a general theoretical description of quantum critical points associated with the disappearance of a Fermi surface.

\section*{Acknowledgments}
I thank Matthew Fisher, B. Halperin, P. A. Lee, M. Randeria, S. Sachdev, R. Shankar, Q. Si, and A. Vishwanath for useful discussions, and Eun-Gook Moon for a careful reading of the manuscript that unearthed several typos. This work was supported by NSF Grant DMR-0705255.

\begin{appendix}
\section{Mean field calculation of electron spectral function at a Mott transition}
Within the mean field theory the electron Greens function ${\cal G}_c(\vec x, \tau)$ is simply the product of the boson and spinon Greens functions:
\begin{equation}
{\cal G}_c(\vec x, \tau) = {\cal G}_b(\vec x, \tau){\cal G}_f(\vec x, \tau)
\end{equation}
The electron spectral function $A_c(\vec K, \omega)$ for real positive frequencies is then given by
\begin{equation}
\label{convolve}
A_c(\vec K, \omega) = \int_{\vec q}\int_0^\omega d\Omega A_b(\vec q, \Omega) A_f\left(\vec K - \vec q, \omega - \Omega\right)
\end{equation}
with $A_{b,f}$ the boson and spinon spectral functions respectively. At the critical point of interest these
take the form
\begin{eqnarray}
A_b(\vec q, \Omega) & = & {\cal A}  \frac{\theta\left(\Omega^2 - c^2q^2\right)}{\left(\Omega^2 - c^2 q^2\right)^{\frac{2-\eta}{2}}} \\
A_f(\vec q, \Omega) & = & \delta\left(\Omega - \epsilon^f_{ \vec q}\right)
\end{eqnarray}
The boson spectral function is that appropriate for bosons at the superfluid insulator transition at fixed integer density. $\eta \approx 0.04$ is the anomalous exponent for the boson field at this $3D$ XY fixed point. ${\cal A}$ is a non-universal prefactor and $c$ is a non-universal velocity. The spinon spectral function is that of free fermions
with dispersion $\epsilon_f$ defined so that $\epsilon_f = 0$ at the Fermi surface. For $\vec K = \left(K_f + k\right)\hat{x}$ with $k$ small, we may write
\begin{equation}
\epsilon^f_{\vec K - \vec q} \approx v_{F0}(k - q_x) + Cq_y^2
\end{equation}
where $v_{F0}$ is the `bare' Fermi velocity, and $C$ is related to the Fermi surface curvature. Putting these into
Eqn. \ref{convolve}, we notice that with the expected $ z= 1$ scaling, for small $|k|, \omega$ the important region of integration involves $|q_x| \sim |q_y| \sim |\Omega| \sim |k|$. Thus we may drop the curvature term $Cq_y^2$ in the fermion dispersion above. It is then straightforward to recast the integral into a scaling form for the spectral function. We find
\begin{equation}
A_c(\vec K, \omega) \sim |k|^{\eta} g\left(\frac{\omega}{v_{F0}|k|}\right)
\end{equation}
with the scaling function $g$ given by
\begin{equation}
g(x) \sim \int_0^x du \frac{\theta\left(u^2 - \lambda^2\left(1+u-x\right)^2\right)}{\left(u^2 - \lambda^2\left(1+u-x\right)\right)^{\frac{1-\eta}{2}}}
\end{equation}
where $\lambda = \frac{c}{v_{F0}}$. In this mean field calculation the scaling function thus depends on the non-universal dimensionless ratio of the boson and fermion velocities. For $x$ small the $\theta$ function in the integral cannot be satisfied and $g(x) = 0$. $g$ first becomes non-zero for $x > \lambda$. For large $x$, $g(x) \sim x^\eta$. Thus the electron spectral function has sharp singularities at a Fermi surface and has a scale invariant form for small deviations away from it.

\end{appendix}

\end{document}